\documentclass[a4paper,fleqn,usenatbib,useAMS]{mnras}
\usepackage{pdfpages}
\pdfoutput=1
\usepackage{graphicx}
\usepackage{lscape}
\usepackage{txfonts}
\usepackage{url}
\usepackage{lastpage}
\usepackage{gensymb}
\usepackage{supertabular}
\usepackage{longtable}
\usepackage{color}
\usepackage{soul}
\usepackage{nicefrac}
\usepackage{lscape}
\usepackage{rotating}

\title[Formation and evolution of ultra-diffuse galaxies]{Ultra-diffuse galaxies outside clusters: clues to their formation and evolution}

\author[Javier Rom\'an \& Ignacio Trujillo]{Javier Rom\'an$^{1,2}$ \thanks{E-mail:jroman@iac.es} and Ignacio Trujillo$^{1,2}$\\
$^{1}$Instituto de Astrof\'{\i}sica de Canarias, c/ V\'{\i}a L\'actea s/n, 
E-38205, La Laguna, Tenerife, Spain\\
$^{2}$Departamento de Astrof\'{\i}sica, Universidad de La Laguna, E-38206, La 
Laguna, Tenerife, Spain}

\begin{document}

\date{}

\pagerange{\pageref{firstpage}--\pageref{lastpage}} \pubyear{2016}

\maketitle

\label{firstpage}

\begin{abstract}

We identify six ultra-diffuse galaxies (UDGs) outside clusters in three nearby isolated groups (0.014 $<$ $z$ $<$ 0.026) using very deep imaging in three different Sloan Digital Sky Survey filters ($g$, $r$ and $i$ bands) from the IAC Stripe82 Legacy Project.  By comparing with the abundance of UDGs in rich galaxy clusters,  we find that the density of UDGs (i.e. their number per unit mass of the host structure where they are located)  decreases towards the most massive systems. This is compatible with a scenario where UDGs are formed preferentially outside clusters. In the periphery (D $>$ 250 kpc) of our three groups, we identify a population of potential UDG progenitors (two of them confirmed spectroscopically). These progenitors have similar masses, shapes and sizes but are bluer, $g-i\sim$ 0.45 (and for this reason brighter, $\mu_{g}(0)<$ 24 mag arcsec$^{-2}$) than traditional UDGs ($g-i\sim$ 0.76). Passive evolution of these progenitors will transform them into regular (i.e. $\mu_{g}(0)$ $>$ 24 mag arcsec$^{-2}$) UDGs after $\sim$6 Gyr. If confirmed, our observations support a scenario where UDGs are old, extended, low surface brightness dwarf galaxies (M$_\star\sim$ 10$^{8}$ M$_{\sun}$)  born in the field, are later processed in groups and, ultimately, infall into galaxy clusters by group accretion.

\end{abstract}


\section{Introduction}

In the last few years, there has been renewed interest in the study of extended and low surface brightness galaxies \citep[][]{1988ApJ...330..634I,1991ApJ...376..404B,1997AJ....114..635D}. Galaxies with $\mu_{g}(0)>$ 24 mag arcsec$^{-2}$ and $R_{e}>$ 1.5 kpc\footnote{It is worth stressing that there is no particular physical motivation behind this observational definition. An effective radius of 1.5 kpc corresponds to $\sim$3.2 arcsec at the distance of the Coma galaxy cluster. This angular size is roughly the pixel size (2.8 arcsec) of the Dragonfly lens array \citep[the telescope used to define this galaxy population;][]{2015ApJ...798L..45V}.} have been coined ultra-diffuse galaxies (UDGs) by \citet{2015ApJ...798L..45V}. These objects have typical stellar masses around 10$^{8}$ M$_{\sun}$ and relatively red colours ($g-i\sim$ 0.8). There has been an intense debate about the ultimate nature of these galaxies. For instance, \citet{2015ApJ...798L..45V,2016ApJ...828L...6V} suggest the intriguing hypothesis that these galaxies could be failed Milky Way-like objects (L$\star$). On the other hand, using their population of globular clusters, \citet{2016ApJ...830...23B} support the idea that these are failed Large Magellanic Cloud-like galaxies \citep[see also,][]{2016ApJ...822L..31P}. Both theoretically and observationally, there is an increasing agreement towards the idea that the vast majority of UDGs are dwarf galaxies \citep[][]{2015MNRAS.452..937Y, 2016ApJ...819L..20B, 2016ApJ...830...23B, 2016MNRAS.459L..51A, 2017MNRAS.466L...1D,2016arXiv161001595A}.

The extremely low surface brightness of these galaxies makes it almost impossible to measure their distance using spectroscopic redshifts (exceptions being DF44 \& VCC1287). For this reason, these galaxies have been explored in galaxy clusters, where the distance to the UDGs is inferred by their proximity to the massive structure \citep[e.g.][]{2015ApJ...807L...2K,2015ApJ...809L..21M,2015ApJ...813L..15M,2016A&A...590A..20V}. This selection effect towards galaxy clusters could affect our understanding of the nature of these galaxies. In fact, not all the known UDGs are in galaxy clusters, as some have been found outside these structures. At least two UDGs are confirmed spectroscopically \citep{2016AJ....151...96M,2017ApJ...836..191T}, and a large number of candidates are shown in \citet{2017MNRAS.468..703R}. These last authors, exploring a wide area of 8 $\times$ 8 Mpc around the galaxy cluster Abell 168, find UDGs both in the cluster and in the large-scale structure surrounding this massive object. Having established the existence of UDGs both inside and outside clusters, the question that arises is whether these objects are formed outside the clusters and are later aggregated to them through the infall of galaxy groups. Answering this question could give us important hints on the formation mechanisms of UDGs. In this paper we explore this scenario.

To achieve our goal, we have probed the presence of UDGs in a number of well known galaxy groups in the deep Stripe 82 survey \cite{2008AJ....135.1057J,2009ApJS..182..543A}. We benefit from the careful reduction of this dataset (the IAC Stripe 82 Legacy Survey) performed by \citet{2016MNRAS.456.1359F}. Among the different groups available in the IAC Stripe82 Legacy Survey, we selected  Hickson Compact Groups (HCGs) \citep{1982ApJ...255..382H}. These are especially useful for our analysis as they are isolated groups by definition. Consequently, the spatial association of UDGs presented in the field of view with these objects is more straightforward.  

This paper is structured as follows. In Section 2, we present the data set and in Section 3, we explain the criteria for selecting UDGs. Section 4 describes how UDGs can be grouped into two differentiated samples according to their colour characteristics and stellar population properties. In Section 5, we present an evolutionary scenario linking the properties of the blue and red populations of UDGs. Section 6 shows how the abundance of UDGs is tightly correlated with the halo mass of the structure where they are embedded. Finally, we discuss our results in Section 7. We adopt the following cosmology ($\Omega_m$ = 0.3, $\Omega_\Lambda$ = 0.7 and H$_0$ = 70 km s$^{-1}$ Mpc$^{-1}$). We use the AB-magnitude system in this work.

\section{Data} 

The images used in this work are based on the IAC Stripe 82 Legacy Survey \citep[][]{2016MNRAS.456.1359F}\footnote{\url{http://www.iac.es/proyecto/stripe82/}}. This survey consists of new deep coadds of the Sloan Digital Sky Survey (SDSS) Stripe82 data, carefully stacked to preserve the faintest surface brightness structures. The pixel scale of these images is the same as regular SDSS data, i.e. 0.396 arcsec. The average seeing of our dataset is around 1 arcsec. In this work, we use the rectified images of the survey. In these images, the residuals of the stacking process have been removed and the sky level has been measured with high precision. This produces high-quality and homogenized deep images.  The mean limiting surface brightness of our data set 
(3$\sigma$; 10$\times$10 arcsec boxes) are 29.1, 28.6 and 28.1 mag arcsec$^{-2}$ for the $g$, $r$ and $i$ bands, respectively. This is $\sim$1.2 mag deeper than the Dragonfly images used to explore UDGs \citep{2015ApJ...798L..45V} and a similar depth like in \citet[][]{2015ApJ...807L...2K}.  

Galaxy groups are gravitationally bound structures with typical values of M$_{200} \sim $ 10$^{13}$ M$_{\odot}$ and R$_{200} \sim$ 500 kpc. The groups we explore here are: HCG07 (RA = 9.816, Dec. = +0.888, $z$ = 0.0141), HCG25 (RA = 50.182, Dec. = -1.052, $z$ = 0.0212) and HCG98 (RA = 358.55, Dec. = +0.37, $z$ = 0.0266). HCGs are defined based on the number of bright galaxy members ($\geq$4) within some specific magnitude range. Together with the number of bright galaxies, HCGs are also defined based on criteria of isolation and compactness, which make them dense galactic associations. These structures can be as dense as the centre of rich clusters, but with modest velocity dispersions. They are expected to be dynamically dominated by  dark matter \citep{1992ApJ...399..353H,2012A&A...539A.106P}. X-ray emission \citep{1996MNRAS.283..690P} and intra-group diffuse light \citep[e.g.][]{2000AJ....120.2355N,2003ApJ...585..739W,2005MNRAS.364.1069D,2008MNRAS.388.1433D,2017MNRAS.464..957H} have been detected in these objects, confirming their spatial association and intense dynamic activity. We assume a spatial scale of 0.288 kpc arcsec$^{-1}$ for HCG07, 0.429 kpc arcsec$^{-1}$ for HCG25  and 0.535 kpc arcsec$^{-1}$ for HCG98.

To explore our groups, we have created wide-field imaging mosaics using the software SWarp \citep{2002ASPC..281..228B}. This allow us to combine different images from the IAC Stripe 82 project and increase the search area. For the HCG07 group, we explored the area 9.5\degree $<$ RA $<$ 10.5\degree and 0.25\degree $<$ Dec. $<$ 1.25\degree equivalent to 1.04 $\times$ 1.04 Mpc at the group distance. For HCG25, the area probed was 49.5\degree $<$ RA $<$ 50.5\degree and -1.25\degree $<$ Dec. $<$ -0.75\degree, i.e. 1.54 $\times$ 0.77 Mpc (in this case the area is limited in declination by the spatial coverage of the Stripe 82 survey). Finally, for HCG98, the area used is 358\degree $<$ RA $<$ 359\degree and -0.25\degree $<$ Dec. $<$ 0.75\degree equivalent to 1.93 $\times$ 1.93 Mpc.

\section{Identification of ultra-diffuse galaxies} \label{sec:Identification}

A first list of galaxy candidates in our mosaics was done using SExtractor \citep[][]{1996A&AS..117..393B}. We require all our sources to be detected simultaneously in the $g$, $r$ and $i$ bands. To remove as much as possible the contamination from point sources, all our targets have a stellarity factor below 0.2 and a minimum area of 15 pixels ($\sim$2.35 arcsec$^{2}$). In addition, we require that the selected sources satisfy the following colour cuts: $g-r$ $<$ 1.4 and $g-i$ $<$ 1.8. These colour criteria remove a large number of targets that have colours not representative of nearby populations (i.e. background contamination). After applying these restrictions, we reduce our initial sample to $\sim$15000 galaxies deg$^{-2}$.

All the sources in the previous sample were fitted using a single S\'ersic model \citep[][]{1968adga.book.....S}. The fitting code used was IMFIT \citep{2015ApJ...799..226E}. The S\'ersic models  were convolved with the point spread function (PSF) of the image. The IAC Stripe 82 Legacy Survey provides a PSF representative of the local conditions of the image. Each piece of the Stripe 82 IAC survey (0.5\degree $\times$ 0.5\degree) has its own PSF. As input parameters for IMFIT, we use the spatial coordinates of the source, the position angle and the effective radius retrieved from the previous SExtractor run. In addition, we mask the closest sources to the target under study. 

To avoid missing potential UDGs sources, we conduct the following sanity checks on the outputs of our model fitting. Any time that IMFIT produces structural parameters representative of a bad fit (like S\'ersic index close to 0 or very large R$_{e}$) or where the magnitude of the model is different from the magnitude obtained from SExtractor by more than 1 mag, then we restart the IMFIT modeling again with a different random input seed and slightly different masking configurations. This process is repeated  until a robust solution is reached. The structural parameters obtained from the IMFIT S\'ersic fit are position angle, ellipticity, S\'ersic index $n$, effective radius (along the semi-major axis) R$_e$ and the global magnitude in each band. 

\begin{figure*}
  \centering
   \includegraphics[width=1.0\textwidth]{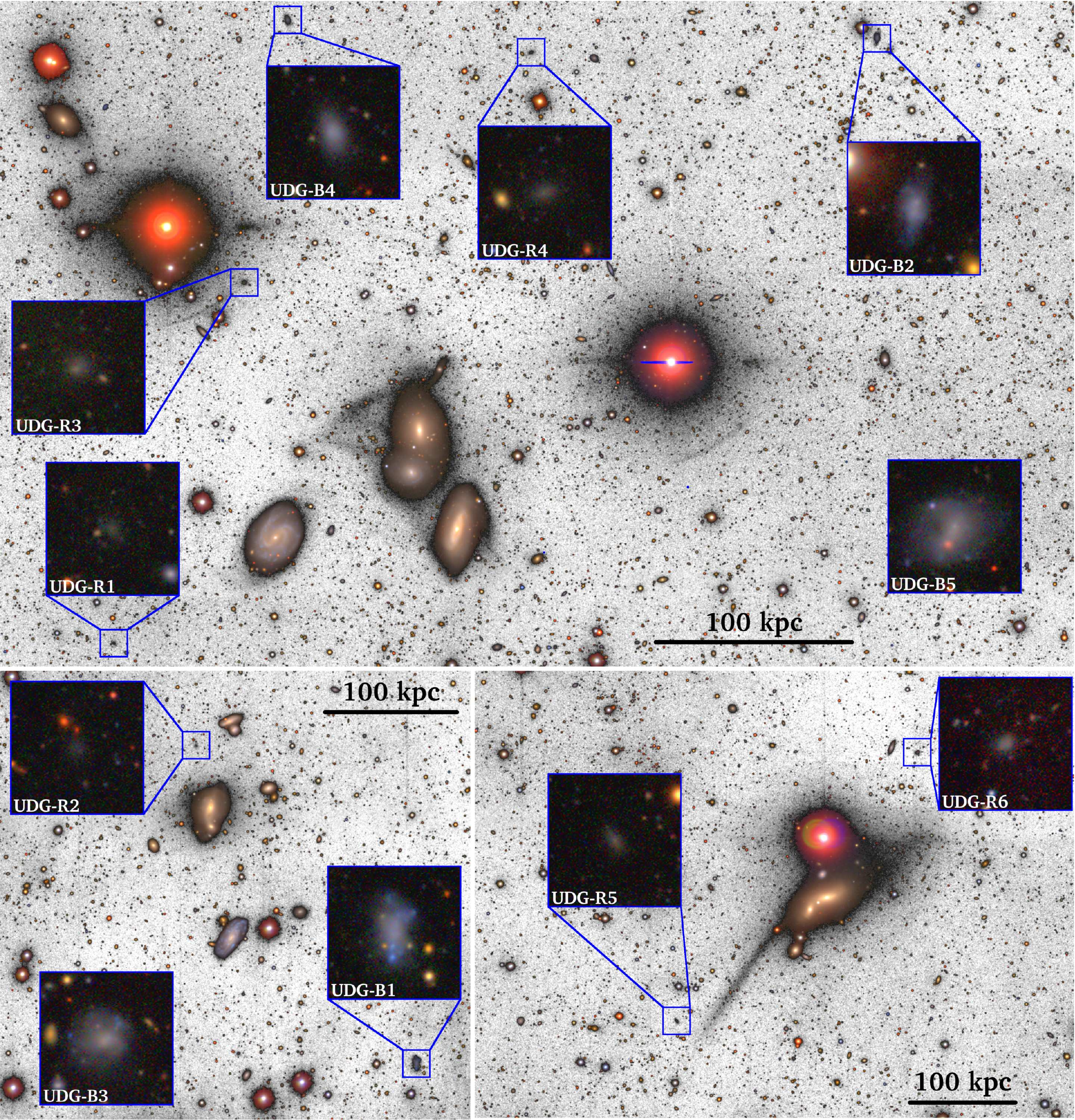}%0.635
    \caption{UDGs discovered around the Hickson compact groups HCG07 (upper panel), HCG25 (lower left panel) and HCG98 (lower right panel). The sizes of these images (which are a portion of our original mosaics) are:  HCG07 (31.65 $\times$ 19.64 arcmin), HCG25 (13.91 $\times$ 13.20 arcmin) and HCG98 (17.66 $\times$ 13.20 arcmin). The inset boxes (50$\times$50 arcsec) are a zoom-in to the individual UDGs. UDG-B5 and UDG-B3 are outside the field of view shown in this figure.}

   \label{fig:HCGs}
\end{figure*}

Once the structural parameters of all galaxies in our data are determined, we select those with R$_{e}>$ 1.3 kpc (in the $g$ band) and observed $\mu_{g}(0)>$ 23.5 mag arcsec$^{-2}$. This provides around 40 UDG candidates. However, an important fraction of these sources are artifacts produced by misidentification of objects in the neighbourhood of bright stars or galaxies. Therefore, we visually explore all the UDG candidates and we reject all the false positives. After this visual inspection, we end up with a final sample of 11 UDGs.  Additionally, for this clean sample, we double check their structural parameters by masking, if necessary, any close or overlapping source to the target that were not masked in the automatic masking process. As a last step, we obtain the extinction values for each galaxy from \citep{2011ApJ...737..103S} using their spatial coordinates and we correct their magnitudes.

\newpage
Note that the final sample of UDGs is located at near projected distances from the host groups, and we are exploring wider areas. These HCGs are isolated structures both in redshift and spatially, consequently, the proximity of the UDGs to these groups shows that any possible contamination by interlopers, if present, must be very low.

\section{Two types of ultra-diffuse galaxies: blue and red populations}

\begin{table*}
\caption{UDGs sorted by increasing $g-i$ colour. All magnitudes and derived parameters are corrected from Galactic extinction.}
\label{tab:params}
\begin{tabular}{ccccccccccc}
\hline
{ID}  &  {RA (\degree)} &  {Dec. (\degree)} &  {$R_{e}$} & {$\mu_{g}(0)$} & {$n$}  & {$b/a$} &   {$g-i$} & {$M_{g}$} & {$M_{*,g-i}$}  & {$D$}\\
  & {(J2000)} & {(J2000)} & {[kpc]} & {[mag arcsec$^{-2}$]} & & & {[mag]} & {[mag]} & {[10$^{8}M_{\sun}$]} & {[kpc]}\\
\hline
UDG-B1 & 50.088 & -1.170 &  3.7 $\pm$ 0.4 &   24.0$^{+0.4} _{-0.5}$  &  0.61 $\pm$ 0.15 & 0.46 $\pm$ 0.06 &  0.27 $\pm$ 0.05 &  -17.01 & 2.2$^{+0.3} _{-0.3}$   & 234\\
UDG-B2 &  9.599 & +1.106 &  2.0 $\pm$ 0.2 &   24.0$^{+0.5} _{-0.6}$  &  0.87 $\pm$ 0.22 & 0.50 $\pm$ 0.07 &  0.41 $\pm$ 0.05 &  -15.23 & 0.6$^{+0.1} _{-0.1}$	& 319\\
UDG-B3 & 49.960 & -0.855 &  3.2 $\pm$ 0.3 &   23.3$^{+0.6} _{-0.7}$  &  1.02 $\pm$ 0.25 & 0.86 $\pm$ 0.12 &  0.50 $\pm$ 0.05 &  -16.70 & 3.4$^{+0.5} _{-0.4}$  & 458\\
UDG-B4 &  9.889 & +1.115 &  1.7 $\pm$ 0.2 &   23.8$^{+0.6} _{-0.7}$  &  1.00 $\pm$ 0.25 & 0.60 $\pm$ 0.09 &  0.53 $\pm$ 0.05 &  -14.95 & 0.7$^{+0.1} _{-0.1}$  & 247\\
UDG-B5 &  9.969 & +0.383 &  3.1 $\pm$ 0.3 &   23.9$^{+0.6} _{-0.6}$  &  0.96 $\pm$ 0.24 & 0.75 $\pm$ 0.11 &  0.55 $\pm$ 0.05 &  -16.10 & 2.2$^{+0.3} _{-0.3}$ & 547\\
\\
UDG-R1 &  9.974 & +0.808 &  1.4 $\pm$ 0.1 &   26.1$^{+0.5} _{-0.6}$  &  0.79 $\pm$ 0.20 & 0.64 $\pm$ 0.09 &  0.63 $\pm$ 0.15 &  -12.46 & 0.1$^{+0.1} _{-0.1}$  & 183\\
UDG-R2 & 50.196 & -1.014 &  1.8 $\pm$ 0.2 &   25.9$^{+0.5} _{-0.5}$  &  0.77 $\pm$ 0.19 & 0.73 $\pm$ 0.10 &  0.66 $\pm$ 0.15 &  -13.31 & 0.2$^{+0.1} _{-0.1}$  & 62 \\
UDG-R3 &  9.910 & +0.985 &  1.5 $\pm$ 0.1 &   25.0$^{+0.6} _{-0.6}$  &  0.92 $\pm$ 0.23 & 0.90 $\pm$ 0.13 &  0.74 $\pm$ 0.15 &  -13.41 & 0.4$^{+0.2} _{-0.1}$  & 140\\
UDG-R4 &  9.769 & +1.099 &  1.8 $\pm$ 0.2 &   25.4$^{+0.6} _{-0.7}$  &  0.98 $\pm$ 0.24 & 0.63 $\pm$ 0.09 &  0.78 $\pm$ 0.15 &  -13.36 & 0.4$^{+0.2} _{-0.1}$  & 224\\
UDG-R5 &358.616 & +0.322 &  2.1 $\pm$ 0.2 &   25.3$^{+0.5} _{-0.6}$  &  0.81 $\pm$ 0.20 & 0.59 $\pm$ 0.08 &  0.85 $\pm$ 0.15 &  -14.21 & 1.1$^{+0.5} _{-0.3}$ & 157\\
UDG-R6 &358.497 & +0.454 &  1.8 $\pm$ 0.2 &   24.1$^{+0.6} _{-0.6}$  &  0.93 $\pm$ 0.23 & 0.73 $\pm$ 0.10 &  0.87 $\pm$ 0.15 &  -14.92 & 2.0$^{+1.0} _{-0.6}$  & 192\\
\hline
\end{tabular}
\end{table*}

\begin{figure*}
  \centering
   \includegraphics[width=0.8\textwidth]{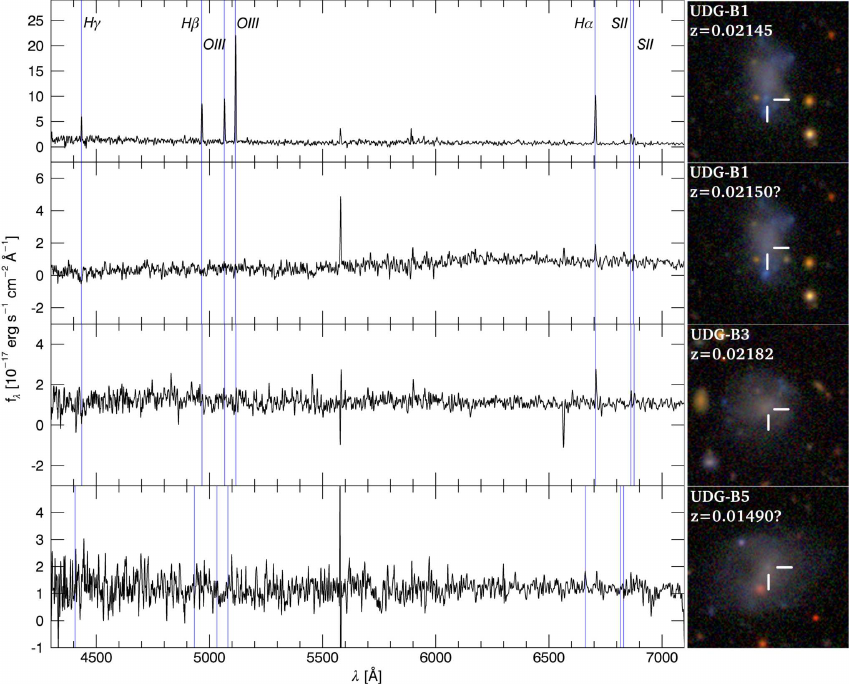}%0.635
    \caption{Available spectroscopic data from the SDSS survey of the UDGs in this work. The redshifts with a question mark have insufficient signal-to-noise ratio for a robust match, and indicate the most likely redshift. The UDG-B5 redshift remains doubtful.}
   \label{fig:specs}
\end{figure*}

\begin{figure}
  \centering
   \includegraphics[width=0.47\textwidth]{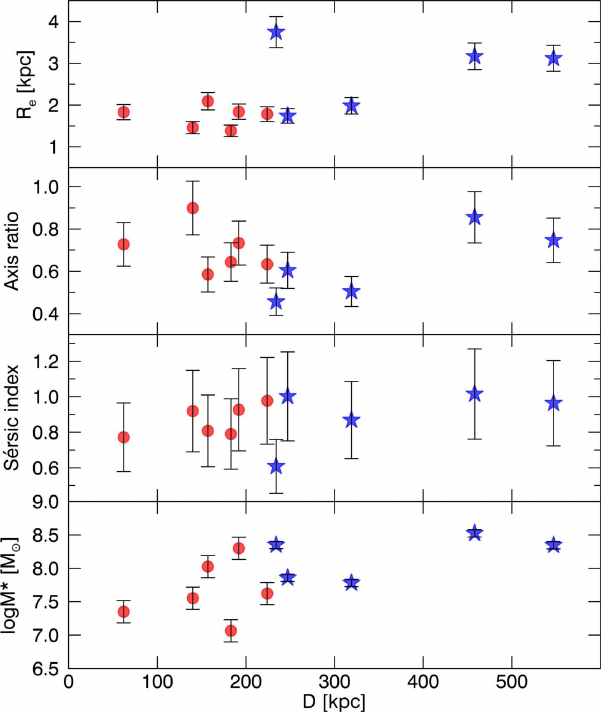}%0.635

    \caption{Structural and stellar mass properties of the UDGs presented in this work. The blue stars represent the population of blue UDGs ($g-i$ $<$ 0.6) and the red dots the red UDGs ($g-i$ $>$ 0.6). Red UDGs are predominantly located at a projected distance less 200 kpc from the group central regions.}

   \label{fig:dist}
\end{figure}

The analysis of the population of extended low surface brightness galaxies in the neighbourhood of our groups shows two types of objects. On the one hand, we find a subset of six galaxies (three in HCG07, one in HCG25 and two in HCG98) with structural and colour properties like those UDGs reported previously in the literature, i.e. R$_{e} \geqslant$ 1.4 kpc, $\mu_{g}(0) \geqslant$ 24.1 mag arcsec$^{-2}$, mean S\'ersic index n = 0.86 $\pm$ 0.04 and mean colour $g-i$ = 0.75 $\pm$ 0.04. These galaxies are at a mean projected distance of $D\approx$ 160 kpc, i.e. quite close to the group centres. On the other hand, there is a number (three in HCG07 and two in HCG25) of extended low surface brightness galaxies in our field of view that are significantly bluer than the previous population. They have 0.27 $< g-i <$ 0.55 and structural parameters R$_{e} \gtrsim$ 1.7 kpc and 23.3 $<\mu_{g}(0) <$ 24.0 mag arcsec$^{-2}$. These objects are located at projected radial distances significantly further away than the redder sample: D $\approx$ 360 kpc. Importantly, three of the five blue UDGs (UDG-B1, UDG-B3, and UDG-B5) have spectroscopic data from SDSS (see Fig. \ref{fig:specs}), two of which have robust redshifts confirming their association with the group structure. The properties of these 11 galaxies are shown in Table \ref{tab:params} and their morphological appearance can be observed in Fig \ref{fig:HCGs}.  We name these galaxies according to their $g-i$ colour as UDG-R (red ones; $g-i>$ 0.6) and UDG-B (blue ones; $g-i<$ 0.6). The relation between the structural properties of the UDGs and the projected distance to the group centre is shown in Fig. \ref{fig:dist}. These results are like those found in \citet[][]{2017MNRAS.468..703R}, with a decrease in stellar mass, the S\'ersic index and the effective radius of the UDGs towards the centre of the group. The morphologies of the blue UDGs are quite irregular, with some of them showing blue clumps of apparently intense star formation. Despite their irregular morphologies, the average S\'ersic index of the blue UDGs is quite similar to the red population, i.e. n $\sim$ 0.9.

\subsection{Stellar population properties of the ultra-diffuse galaxies}

The very low surface brightness of the UDGs makes studying their stellar population properties using spectroscopy with present-day instrumentation extremely challenging. Some of our brightest galaxies have a spectra provided by the SDSS. However, their signal-to-noise ratio is insufficient to get reliable information about their stellar population properties. There is one exception, for the UDG-B1 galaxy, in which the SDSS pipeline took a spectrum of one of its bright blue clumps located at the spatial position (RA = 50.088, Dec. = -1.172). According to that spectrum, this region of the galaxy has a low metallicity and extremely young age ($<$ 0.1 Gyr). Due to the colour characteristics of this clump (significantly much bluer than the rest of the object), we consider that these stellar population properties are not representative of the whole galaxy. For the above reasons, we need to address the problem of the stellar population properties of the UDGs using integrated deep photometry. In this paper, we use the $g$, $r$ and $i$ bands for this. Some of our galaxies, especially the bluer ones, are detected in the $u$ band as well. However, we restrict ourselves to only those SDSS filters where all the galaxies in the sample have been detected. 

\begin{figure*}
  \centering
   \includegraphics[width=\textwidth]{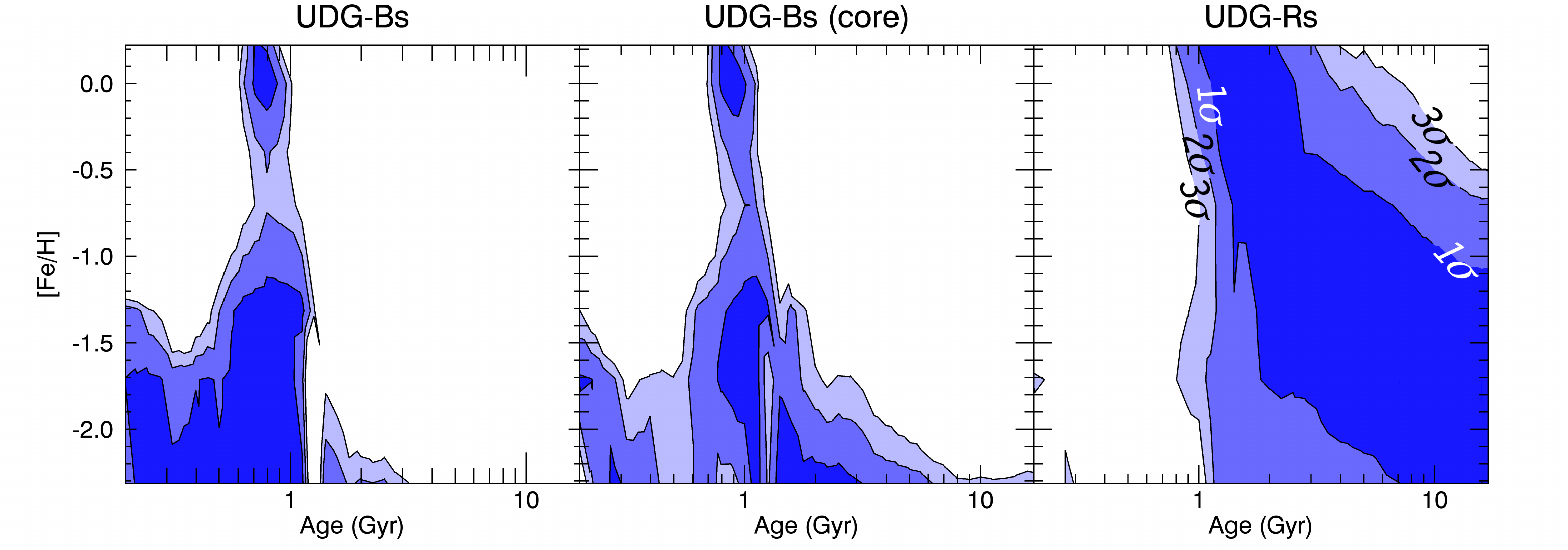}%0.635
    \caption{Age and metallicity distribution of our sample of UDGs. The different colour regions correspond to the 1$\sigma$, 2$\sigma$ and 3$\sigma$ confidence intervals.}
   \label{fig:agemet}
\end{figure*}

To have a rough estimation of the stellar masses of our galaxies, we use the method provided by \citet{2015MNRAS.452.3209R} (assuming a Chabrier initial mass function). In particular, we use the $g-i$ colour and the absolute magnitude in the $r$ band. The stellar masses are provided in Table \ref{tab:params}. The mean stellar mass of the blue UDGs is $\sim$1.8 $\times$ 10$^{8}$ M$_{\odot}$, whereas the red ones are a factor of $\sim$2 less massive: $\sim$0.9 $\times$ 10$^{8}$ M$_{\odot}$. As a further test, we have re-estimated the stellar masses of our UDGs using the stellar population predictions provided by \citep{2015MNRAS.449.1177V} using a universal Kroupa IMF \citep{2001MNRAS.322..231K}. We use the $g-r$ and $r-i$ colour map as proxy for estimating the best age and metallicities describing the observed colours. We follow a similar approach as the one used in \citet{2014ApJ...794..137M}. The outcome of this exercise is given in Fig. \ref{fig:agemet}. Using this new method, the mean stellar mass of the blue UDGs is $\sim$1.2 $\times$ 10$^{8}$ M$_{\odot}$, whereas the red UDGs have a mean stellar mass of $\sim$0.6 $\times$ 10$^{8}$ M$_{\odot}$. Both approaches provide similar stellar masses for the two populations. The clumpy appearance of the blue UDGs suggests that the colours (and consequently, the stellar population properties) of these galaxies can be heavily affected by these knots of intense star formation. In this sense, it is worth exploring whether the stellar masses of the blue UDGs would change on using the colours of the central part (R $<$ 4 arcsec)\footnote{Using a R $<$ 4 arcsec aperture, we get to select the core regions for all galaxies avoiding blue knots.} of these objects, where the star formation is less prominent. Using these colours we estimate the ($M/L$)$_i$ and together with the absolute magnitude of the galaxies in the $i$ band (to minimize the contribution of the star forming regions) we derive again the stellar masses of the blue population. We obtain a mean stellar mass for the blue UDGs of $\sim$1.4 $\times$ 10$^{8}$ M$_{\odot}$, in good agreement with all the previous different methods.

Once the stellar masses of our UDGs were estimated, we focussed our attention on the ages and metallicities of the two populations of UDGs. Having only two colours, the expected degeneracy between the age and the metallicity is very high. This is, in fact, what we see in Fig. \ref{fig:agemet}.  The red UDGs  show mean colours of $g-r$ = 0.55 $\pm$ 0.04 and $g-i$ = 0.75 $\pm$ 0.04 ($r-i$ = 0.20 $\pm$ 0.04), compatible with a metallicity in the range -2 $<$ [$Fe/H$] $<$ -1  and $t$(Gyr) $>$ 2 Gyr, although the possibility that red UDGs have a younger population 1 $<$ $t$(Gyr) $<$ 2 Gyr with a solar metallicity is, however, not rejected. These metallicities and age ranges are in agreement with previous works. For instance, \citet[][]{2015ApJ...798L..45V} find $g-i$ = 0.8, suggesting a stellar population with 7 Gyr and [$Fe/H$] = -1.4 or 4 Gyr and [$Fe/H$] = -0.8. Other authors, like  \citet[][]{2016A&A...590A..20V} find $g-r$ = 0.6 populations with an age of 2 Gyr assuming solar metallicity or 6 Gyr with  [$Fe/H$] = -0.7. Note that the gap present around 1.2 Gyr in the age-metallicity map corresponds to the transition from the light contribution of AGB to RGB stars \citep[see e.g.][]{1994A&AS..106..275B}.

Finally, the sample of blue UDGs present the following mean colours: $g-r$ = 0.30 $\pm$ 0.03 and $r-i$ = 0.15 $\pm$ 0.03. These bluer colours imply that $t$(Gyr) $<$ 1 Gyr (see Fig. \ref{fig:agemet}). However, we can say little about their metallicities. If we focus our attention to the core of the blue UDGs, their redder colours $g-r$ = 0.34 $\pm$ 0.04 and $r-i$ = 0.17 $\pm$ 0.02 are suggestive of a slightly older population with 1 $\lesssim$ $t$(Gyr) $\lesssim$ 2 Gyr.

\begin{figure*}
  \centering
   \includegraphics[width=1.0\textwidth]{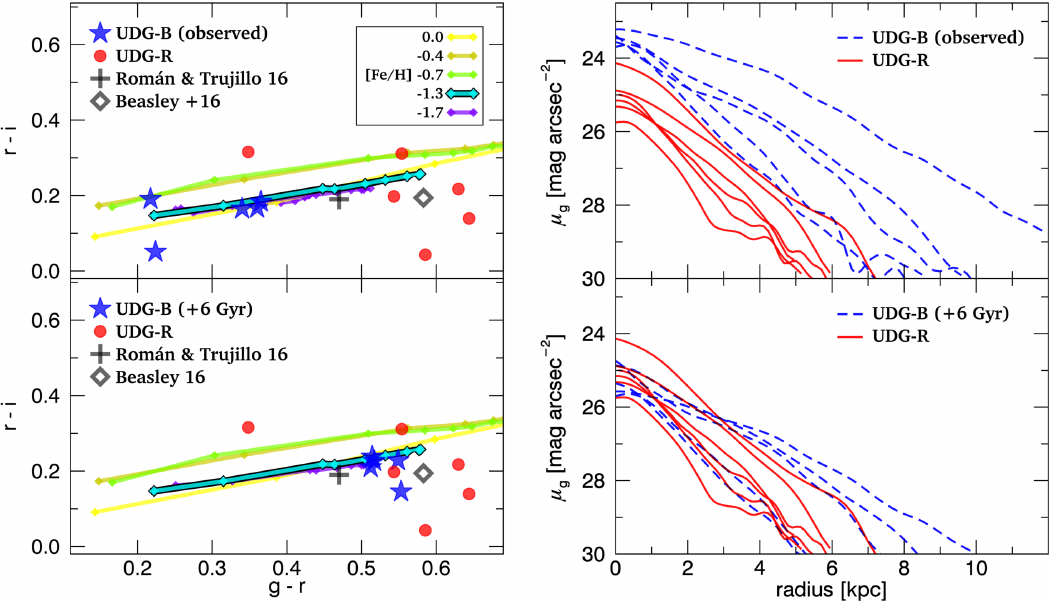}%0.635
   
    \caption{An evolutionary scenario between blue and red UDGs. \textbf{Left column}: colour-colour map for the UDGs in this work. The upper panel shows the location of the blue UDGs (blue stars) and red UDGs (red dots). Overplotted are the stellar population tracks from the \citet{2015MNRAS.449.1177V} models. We also show the location of the UDGs from the work by \citet{2017MNRAS.468..703R} and \citet{2016ApJ...819L..20B}. The ticks on the stellar population tracks represent the ages (from left to right): 0.5, 1.0, 2.0, 3.2, 5.0, 7.0, 10.0, 12.5 and 15 Gyr. The lower panel shows the location of the blue UDGs after 6 Gyr of passive evolution. \textbf{Right column}: The upper  panel shows the observed surface brightness profiles in the $g$ band of the blue and red UDGs. The bottom panel shows the expected location of the surface brightness profiles of the blue galaxies after 6 Gyr of passive evolution (there has been no attempt to model any mass or size evolution).}

   \label{fig:profiles}
\end{figure*}

\section{An evolutionary scenario}

In this section, we explore the following evolutionary scenario: red UDGs are the result of the group environmental processing of the blue UDGs located in the outskirts. In other words, blue UDGs eventually infall into the group centre where their gas is removed by ram pressure or tidal stripping and they become red UDGs with time. In this process, their stellar population gets redder and older but their metallicity should not change. During this process, it is likely that the UDGs lose part of their stellar mass (decreasing their effective radius) and become more rounded. Is this evolutionary scenario compatible with our observations? To start with, blue UDGs are, in fact, systematically located in the outskirts of the groups (see Fig. \ref{fig:dist}). In that sense, the star formation activity of the UDGs seems to be strongly related with the location of these galaxies within the dark matter halo of the group. Moreover, the sizes and stellar masses of the red UDGs are smaller than the blue UDGs. The red UDGs are also roundish.

We can try to quantify more the above evolutionary scenario more by comparing the stellar population properties of blue and red UDGs with the expected theoretical transformation for dwarf galaxies within group environments. \citet{2015MNRAS.453...14Y} studied the environmental processing of dwarf galaxies  infalling into groups (10$^{13}$ M$_{\odot}$). According to these authors, the dwarf galaxies become gas poor after 6 Gyr of the first infall. Note, however, that these simulations are for dwarf galaxies that are more massive (10$^{9}$ M$_{\odot}$) than our blue UDGs (10$^{8}$ M$_{\odot}$). Having said that, it is reasonable then to explore whether a passive evolution of our blue UDGs will lead to properties that resemble those of our red UDGs.

We conduct such an exercise in Fig. \ref{fig:profiles}. In this figure (upper left panel) we show the location of our UDGs (red and blue) in the colour-colour map $g-r$ versus $r-i$. Overplotted on the galaxies are the tracks (for different metallicities) of the time evolution of passively evolving stellar populations from \citet{2015MNRAS.449.1177V} models. To simplify our exercise, we calculate the passive evolution of the colours of the blue UDGs for a fixed metallicity. We select [$Fe/H$] = -1.31\footnote{A solar metallicity will also represent the colour position of the blue UDGs but considering the low mass of our UDGs we prefer to do this task assuming they have a low metallicity. Low mass galaxies are not expected to have solar metallicity because of the inefficiency of their star formation \citep[e.g.][]{1996ApJS..106..307V}.}. Once we have selected a metallicity, it is straightforward to add a colour increase to the observed colours of the blue UDGs after a given amount of time. In our case, we decided to use 6 Gyr (although our main results are basically unchanged if the time evolution is selected to be within 6 to 10 Gyr)\footnote{Every Gyr increase produces a reddening in our colours of $\sim$0.01 in this time interval.}. We use 6 Gyr motivated by the simulation of \citet{2015MNRAS.453...14Y}. The results of the time evolution of the blue UDGs in the colour-colour map are shown in the  lower left panel of Fig. \ref{fig:profiles}. The location on this map of the average UDG in the Abell 168 cluster and its surrounding is also shown \citep{2017MNRAS.468..703R}. In addition, we include also the location of the galaxy VCC1287 \citep[][]{2016ApJ...819L..20B}. We add these extra points as these are the only studies (besides this) with a characterization of UDGs in three optical bands. The scatter of the red UDGs in this colour-colour map is larger due to the higher photometric uncertainties at measuring the colours of these fainter objects. Note how, after 6 Gyr of passive evolution, the location of the blue UDGs are in nice agreement with the location of the red UDGs. It is important to stress that some of our galaxies (like UDG-B4) have some properties in between the red and the blue population, reinforcing the idea that both populations are connected.

\begin{figure*}
  \centering
   \includegraphics[width=0.9\textwidth]{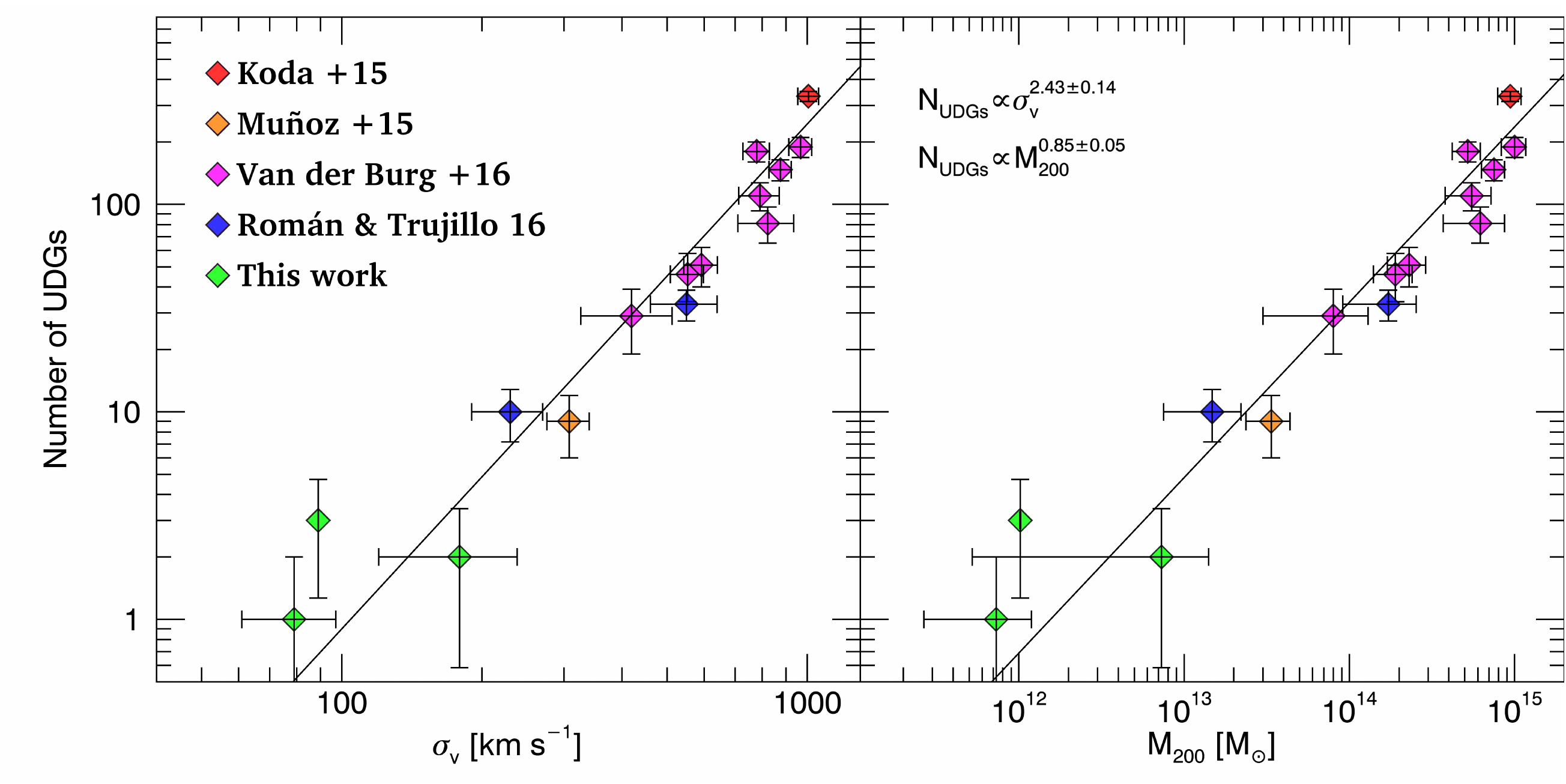}%0.635
    \caption{The abundance of UDGs as a function of the velocity dispersion and halo mass of the structure where they are located. The data shown is a compilation from the literature and our own work here. See text for further details.}
   \label{fig:sigmas}
\end{figure*}

Once the evolution in the colour-colour map of the blue UDGs has been explored, it is worth looking at the change of the surface brightness profiles of these objects assuming a passive evolution. This is, of course, an oversimplification of the real scenario. In fact, it is expected that these infalling dwarf galaxies will have some stellar mass loss during the infall process to the group. An eventual stellar mass loss will produce a decrease in their effective radius (as most of the mass loss will be likely produced in the outskirts of these objects). The right column (upper panel) in Fig. \ref{fig:profiles} shows the observed surface brightness profiles of both blue and red UDGs. The passive evolution of the blue UDGs is modeled in the lower panel of the same figure. As can be seen, after 6 Gyr of passive evolution all the current blue UDGs will be classified as regular UDGs (i.e. large objects with $\mu_{g}(0)$ $>$ 24 mag arcsec$^{-2}$). As the S\'ersic index of both blue and red UDGs are very similar (n $>$ 0.8), a passive transformation of the blue population will resemble the red population. However, despite the similar appearance of the evolved blue UDGs in relation to the red UDGS, some of the evolved blue UDGs show an excess of light in their outer regions compared to the red population. This is as expected considering that: (i) we have not modeled any mass loss in the profiles of the blue UDGs and (ii) the profiles of the blue UDGs are affected by the presence of intense star-forming regions in the outer parts. Considering that these star-forming regions are relevant in light but not as much in mass, the time evolution of the surface brightness profiles of the blue galaxies in their outer regions will approximate the shape of the red population. Summarizing, under the hypothesis that the blue UDGs will passively evolve as the result of their infall into the group's gravitational potential, the global colour and shape properties will be close to the red UDG population, making a scenario linking the two type galaxies plausible.

\section{The abundance of UDGs as a function of the halo mass}

\citet{2016A&A...590A..20V} showed that there is a tight correlation between the number of UDGs in a given cluster and the mass of the cluster as parametrized by M$_{200}$. According to that work, the abundance of UDGs increases almost proportionally to the mass of the cluster in which they are embedded. \citet{2016A&A...590A..20V} studied this relation for clusters with M$_{200}$ $\gtrsim$10$^{14}$M$_{\odot}$. In this work, we want to probe where that relation also holds for less massive gravitationally bound systems. To do this, we have compiled in Fig. \ref{fig:sigmas} the number of UDGs observed in different works as a function of their host velocity dispersion and M$_{200}$. For the clusters provided by \citet{2016A&A...590A..20V}, we use their M$_{200}$ and the velocity dispersion from \citet{2015A&A...575A..48S}. For the Abell 168 cluster and the UGC 842 group presented in \citet{2017MNRAS.468..703R}, we use as sources for their velocity dispersions: \citet{2004ApJ...600..141Y} (for the cluster) and \citet{2010AJ....139..216L} (for the group). For the Coma cluster \citep{2015ApJ...807L...2K}, the velocity dispersion was taken from \citet{1993ApJ...404...38G}, whereas for the Fornax cluster \citep{2015ApJ...813L..15M} the velocity dispersion is obtained from \citet{2001ApJ...548L.139D}.  Finally, for the groups presented in this paper, we have used the velocity dispersions given by  \citet{2006A&A...456..839T}. Once we have the velocity dispersions, we follow \citet{2013MNRAS.430.2638M} to get M$_{200}$ for these structures.

Fig. \ref{fig:sigmas} shows that the tight correlation between the abundance of UDGs and the halo mass of the structures in which they are located extend also towards lower masses than those originally explored by \citet{2016A&A...590A..20V}. We have fitted our relations assuming a power law, as was done by \citet{2016A&A...590A..20V}: N $\propto$ $\sigma^{\alpha}$ and N $\propto$ M$_{200}^{\beta}$. We find the following values: $\alpha$ = 2.43 $\pm$ 0.14 and $\beta$ = 0.85 $\pm$ 0.05. In our plot of M$_{200}$, the estimation of this quantity comes from two different sources \citep{2015A&A...575A..48S, 2013MNRAS.430.2638M}, however, our results remain basically the same ($\beta$ = 0.87 $\pm$ 0.05) if we use the methodology of \citet{2013MNRAS.430.2638M} for estimating M$_{200}$ for all the structures. Interestingly, \citet{2016A&A...590A..20V} found a value for $\beta$ = 0.93 $\pm$ 0.16  in agreement (within the error bars) with the expanded mass sample we have explored here. Determining the exponent $\beta$ is key for understanding in which structures the UDGs are formed with higher efficiency. We discuss the implication of our finding in the next section.

\section{Discussion and Summary}

In this paper, we have suggested an evolutionary scenario where present-day UDGs are the result of the environmental transformation (mainly by gas stripping) of infalling low surface brightness dwarf galaxies (of similar stellar mass and structural parameters) into the gravitational potential of the groups and clusters. We want to expand such discussion further and explore which kind of structures (groups or clusters) are more favorable for the formation of UDGs. Initially, clusters were considered the natural place for finding UDGs; however, there is increasing evidence (first noted by \citet{2017MNRAS.468..703R} and recently by \citet{2016ApJ...833..168M}; see also: \citet{2016MNRAS.463.1284O,2016A&A...596A..23S}) that UDGs are also found outside clusters. The existence of UDGs in the field has also been motivated theoretically by \citet{2017MNRAS.466L...1D}. As there is a tight correlation between the abundance of UDGs and the mass of the host structures in which they are located \citep{2016A&A...590A..20V}, it is worth exploring what can be learned from this relation.

The relation between the abundance of UDGs and M$_{200}$ is characterized by a power-law with an exponent slightly lower than 1, i.e. $\beta$ = 0.85 $\pm$ 0.05. If $\beta$ were larger than 1, the UDGs would be formed preferentially in clusters. Let us expand on this. Under the assumption \citep[see e.g.][]{2002ASSL..272...79B,2002ASSL..272...39G} that clusters were purely the result of merging of smaller sub-units (i.e. groups of galaxies), the abundance of UDGs will be never larger (i.e. $\beta$ $>$ 1) than the contribution of UDGs accreted through the infalling of groups to the cluster. Consequently, observations showing that $\beta$ $>$ 1 would be a strong argument favouring the preferential (in-situ) formation of UDGs in clusters of galaxies. If $\beta$ $<$ 1, however, the observational result is not straightforward to interpret. Naively, one could understand that if $\beta$ $<$ 1 then UDGs form preferentially in groups. As groups infall into clusters, one would expect that some UDGs could be disrupted during the accretion process and thus, $\beta$ $\leq$ 1. $\beta$ $<$ 1 could be also expected if UDGs are more easily destroyed in clusters than in groups over time\footnote{Note, however, that from the theoretical point view, the situation is the opposite, groups are more effective at destroying dwarf galaxies than clusters \citep[see e.g.][]{2003astro.ph..5512M}.}. For this reason, if $\beta$ $<$ 1, we cannot firmly conclude that UDGs are form preferentially in groups than in clusters. Nonetheless, the observational result does not contradict this hypothesis.

Another interesting issue to explore is why the relation between  the abundance of UDGs and $\sigma$ is so tight (Pearson correlation coefficient $r$ = 0.964). This is not at all expected as the number of UDGs in a given structure is a function of the depth of the survey used to detect these objects and the background and foreground contamination. The limiting surface brightness of the different surveys presented in Fig. \ref{fig:sigmas} are: 29.2 mag arcsec$^{-2}$ (3$\sigma$, 10$\times$10 arcsec in $g$ band; \citet{2017MNRAS.468..703R}), 28.8-29.2 mag arcsec$^{-2}$ (3$\sigma$, 10$\times$10 arcsec in $R$ band; \citet{2015ApJ...807L...2K}),  28.9 mag arcsec$^{-2}$ (3$\sigma$, 10$\times$10 arcsec in $i$ band; \citet{2015ApJ...813L..15M}), 28.9 mag arcsec$^{-2}$ (3$\sigma$, 10$\times$10 arcsec in $r$ band; \citet{2016A&A...590A..20V}). As can be seen, all the surveys have a relatively similar surface brightness limit. This is as expected due to technical limits of present-day telescopes (see a discussion in \citet{2016ApJ...823..123T}). This could explain why the number of UDGs detected is similar among clusters of similar masses for different groups. Another crucial issue is related to the expected contamination by foreground and background interlopers. All the previous authors applied some particular recipe for cleaning their number counts. The tightness of the correlation indicates that the abundance of UDGs is somehow robust to the different methodologies.

To conclude, our observations support a scenario where present-day UDGs are old ($>$2 Gyr), extended, low surface brightness ($\mu$(g,0) $>$ 24 mag arcsec$^{-2}$) dwarf (M$_\star$ $\sim$ 10$^{8}$ M$\sun$) galaxies. These galaxies would have been formed preferentially in the field where they would be brighter (i.e., $\mu$(g,0) $<$ 24 mag arcsec$^{-2}$ and eluding the current criteria for selecting UDGs) and younger.  Later on they would be processed in groups (on scales of around 6 Gyr) and, ultimately, infalled into galaxy clusters by groups accretion.

\section*{Acknowledgments}

We thank the referee for a careful review and useful comments. We want to thank Alejandro Borlaff for his help on producing some of the plot of this paper. We thank also Alejandro Vazdekis, Mike Beasley and Jorge S\'anchez Almeida for helpful discussions. This research was supported by the Instituto de Astrof\'isica de Canarias. The authors of this paper acknowledges support from grant AYA2016-77237-C3-1-P from the Spanish Ministry of Economy and Competitiveness (MINECO). JR thanks MINECO for financing his PhD through FPI grant.

\appendix

\end{document}